\documentclass[12pt,draftclsnofoot,onecolumn]{IEEEtran}
\usepackage[T1]{fontenc}
\usepackage[latin9]{inputenc}
\usepackage{cite}
\usepackage{amsmath,setspace}
\usepackage{amssymb}%
\usepackage{graphicx}
\usepackage{color,upgreek}
\usepackage{algorithm,algorithmic}

\renewcommand{\vec}[1]{\bf{#1}}     
\newcommand{\herm}{^H}
\newcommand{\trans}{^{\mbox{\scriptsize T}}}

\begin{document}
\title{\LARGE A Minorization-Maximization Method for Optimizing Sum Rate in Non-Orthogonal Multiple Access Systems}
\author{Muhammad Fainan Hanif, Zhiguo Ding~\IEEEmembership{Member,~IEEE}, Tharmalingam Ratnarajah~\IEEEmembership{Senior Member,~IEEE}, and
George K. Karagiannidis~\IEEEmembership{Fellow,~IEEE}\thanks{M. F. Hanif and Z. Ding are with the
School of Computing and Communications, Lancaster University, Lancaster LA1 4WA, United Kingdom.
Email: \{m.f.hanif, z.ding\}@lancaster.ac.uk and mfh21@uclive.ac.nz.\par T. Ratnarajah is with the Institute for Digital Communications, School of
Engineering, University of Edinburgh, Edinburgh EH9 3JL, United Kingdom. Email: T.Ratnarajah@ed.ac.uk.\par G. K. Karagiannidis is with
the Department of Electrical and Computer Engineering, Aristotle University
of Thessaloniki, 54 124, Thessaloniki, Greece and with the Department of Electrical and Computer
Engineering, Khalifa University, PO Box 127788, Abu Dhabi, UAE. Email: geokarag@ieee.org.}
}

\maketitle
\begin{abstract}
Non-orthogonal multiple access (NOMA) systems have the potential to deliver higher system throughput,
compared to contemporary orthogonal multiple access techniques. For a linearly precoded multiple-input multiple-output (MISO) system,
we study the downlink sum rate maximization problem, when the NOMA principles are applied. Being a non-convex and intractable optimization problem,
we resort to approximate it with a minorization-maximization algorithm (MMA), which is a widely used tool in statistics. In each step of the MMA, we solve a second-order cone program,
such that the feasibility
set in each step contains that of the previous one, and is always guaranteed to be a subset of the feasibility
set of the original problem. It should be noted that the algorithm takes a few iterations to converge. Furthermore, we study the conditions under which the
achievable rates maximization can be further simplified to a low complexity design problem, and we compute the probability of occurrence of this event. Numerical
examples are conducted to show a comparison of the proposed approach against conventional multiple access systems. NOMA is reported to provide better
spectral and power efficiency with a polynomial time computational complexity.
\end{abstract}
\section{Introduction}
Efficient multiple access techniques in wireless systems has long been a sought after desirable feature. Several facets haven been considered, while dealing
with the design of multiple access schemes. For example, spectral efficiency, reliability and quality of service, efficient utilization of radio resources,
and recently, energy efficiency are some of the objectives, that form the basis of multiple access techniques in wireless communication systems. Non-orthogonal multiple access
(NOMA) has been conceived as a breakthrough technology for fifth generation ($5$G) wireless systems \cite{DOCOMOnoma,Saito,Benje}. The main themes of $5$G networks, namely, reduced latency,
high connectivity, and ultra-fast speeds are being attributed to devising systems working on the principles of NOMA \cite{DOCOMOnoma}. NOMA uses power domain to multiplex additional users
in the time/frequency/code, slot already occupied by a mobile device. The enabling techniques for NOMA are not new and find their roots in some old principles--\emph{superposition coding}
(SC) and \emph{successive interference cancellation} (SIC). SC was first proposed by Cover in \cite{CoverSup}, as an achieveability scheme for a degraded broadcast channel. Likewise, various
versions of SIC have been employed in the past in systems like Vertical-Bell Laboratories Layered Space-Time (V-BLAST) and Code Division Multiple
Access (CDMA) \cite{VBLAST,CDMASIC}. Therefore, in addition to being a candidate for the next generation of
$5$G wireless networks, it is very important that NOMA has also the potential to integrate well with existing multiple access paradigms.\par In NOMA, the base station (BS) transmits a superposition coded signal, which
is a sum of all messages of the users. The users are arranged with respect to their effective channel gains i.e., the one with the lowest gain is assumed to be
at the bottom of the sequence, the one with the highest gain at the top, while the remaining are
arranged in an increasing order between the two. NOMA ensures that the weaker users receive a higher fraction of the total power budget. When a stronger user
is allowed to access the slot being occupied by a weaker one, its signal does not adversely impact the performance of the weaker user,
as it is already experiencing a channel fade. At the same time, the stronger user can get rid of the interference due to the weaker one, by applying
a SIC operation. In traditional orthogonal multiple access schemes, once the slot has been reserved for a user, other users are prohibited from
accessing that. This, of course, has a negative impact on the aggregate systems's throughput. The major outcome of sharing the same channel slot is that the sum rates are
expected to improve, and with intelligent power allocation the weaker users can also be efficiently served.\par \subsection{Literature}To the best of our knowledge, as of today, NOMA has mostly been
explored for single-input single-output (SISO) systems. For example, in \cite{Ding1} Ding {\it et al.} studied NOMA for the downlink of a cellular system, and by assuming fixed powers, they
derived expressions for the aggregate ergodic sum rate and outage probability of a particular user. Interestingly, in that paper it was concluded that in the absence of
a judiciously chosen target data rate, a user can always be in outage. For multiple-input multiple-output (MIMO) systems, Lan {\it et al.} \cite{LanNOMA}, explored the impact of
error propagation of SIC and user velocity on the NOMA performance. Their results showed that even in the worst error propagation scenario, NOMA outperforms
conventional orthogonal multiple access and can yield performance gains for different user mobility. Chen {\it et al.} \cite{ChenNOMA}, studied NOMA for the downlink of a wireless system,
when BS and receivers
are each equipped with two antennas. Traditional minimum-mean-squared-error (MMSE) precoding matrices have been used, which do not guarantee maximum throughput
for a given user ordering. 
Similarly, Timotheou {\it et al.} \cite{Timotheou}, studied the power allocation  for NOMA in a SISO system from a fairness point of view. Finally, Ding {\it et al.}, investigated
MIMO-NOMA in \cite{Ding2}, and derived outage probabilities for fixed and more sophisticated power allocation schemes.\par
\subsection{Contributions} In this paper, we focus on the downlink of a multiple-input single-output (MISO) system, in which the transmit signals of each user are
multiplied by a complex precoding vector. The goal is to design these vectors in order to maximize the total throughput of the system,
while simultaneously satisfying the NOMA constraints. To solve this problem we rely on the approximation technique that has been commonly dubbed as
{\it concave-convex procedure} (CCP)\footnote{If the original problem is a
minimization instead of a maximization, the procedure has been referred to as convex-concave procedure (CCP).} or {\it minorization-maximization algorithm} (MMA)\footnote{The MMA has also been called as
majorization-minimization algorithm if the original problem is a minimization problem.} \cite{MMATut,MMA1,MMA2,CCPTut,CCP1,MMAStoica}. Under the different name of sequential convex programming
a parametric approach has been proposed in \cite{amir}. Recently, in the context of weighted sum rate maximization and
physical layer multicasting, similar ideas were used by Hanif {\it et al.} and Tran {\it et al.} in \cite{HanifTSP,TranSPL}, respectively. 
Due to the flexible nature of MMA approach, these ideas have also
be used in image  processing applications \cite{MMA3}.\par
The main contributions of this paper can be summarized as follows:
\begin{itemize}
\item By incorporating decodability constraints to ensure that better users can perform SIC, we provide a novel mathematical programming based approach to solve the sum rate maximization problem in
the downlink of a MISO system, relying on NOMA principles. Similarly, constraints are also included to guarantee that the desired signals of the
weaker users are strong enough to render them non-zero data rates.
\item Using the MMA concept, we develop an iterative algorithm that solves the NOMA sum rate maximization problem and obtains
complex precoding vectors, which maximize the aggregate throughput. Unlike traditional approaches that rely on semidefinite programming (SDP), to deal with such optimization problems,
the MMA based
algorithm solves a second-order cone program (SOCP) in each step.
\item We show that the proposed algorithm is provably convergent in few iterations. Moreover, a complexity analysis is also carried out to show that the worst case complexity of the
SOCP, which we solve in each run, is just polynomial in design dimensions. Furthermore, under plausible assumptions, the algorithm converges to
the Karush-Kuhn-Tucker (KKT) point of the original problem.
\item We present an approximation to the original optimization program, with the main goal of complexity reduction. To provide more insight, we study conditions
under which this approximation is tight. Moreover, for the special case of orthogonal precoding vectors, we provide a probabilistic insight regarding the tightness of the
proposed approximation.
\item Finally, numerical examples are presented to show the validity of the proposed algorithm. These results reveal that the NOMA transmission outperforms the
conventional orthogonal multiple access schemes, particularly when the transmit signal-to-noise ratio (SNR) is low, and the number of users
are greater than the number of BS antennas. We also investigate the scenario, where
the proposed approximation exactly matches the original problem. In this case, it is shown that the distance between the users and the BS plays a crucial role and affects
the system's throughput.
\end{itemize}
\subsection{Structure}
The rest of the paper is organized as follows. In Section \ref{sysetup}, we describe the system model and formulate the problem. In Section \ref{prereq}, we present the
preliminaries, needed to outline the algorithm in the next section. The algorithm is developed and analysed in Section \ref{proSol}, while a reduced complexity approximation
is motivated and developed
in Section \ref{reDcomp}. Finally, numerical results and conclusions are presented in Sections \ref{NumRes} and \ref{Conclusion}, respectively.\par
\subsection{Notations} Bold uppercase and lowercase letters are used to denote matrices and vectors, respectively. The symbols $\mathbb{C}^n,\mathbb{R}^n$ and $\mathbb{R}^n_+$ are used
for $n$-dimensional complex, real, and nonnegative real spaces, respectively. For a vector ${\vec e}$, its $j^{\textrm{th}}$ coordinate is denoted
by $e_i$. Furthermore, $\|{\vec e}\|_2$ is used to represent $l_2$ norm of a vector ${\vec e}\in\mathbb{C}^n$, which
is defined as $\|{\vec e}\|_2=\sum_{i=1}^n|e_i|^2$, where $|e_i|$ is the absolute value of $e_i$. $\mathcal{O}(.)$ is reserved for complexity estimates. Unless otherwise specified,
calligraphic symbols are used to represent sets. $\lceil x \rceil$ is the ceiling function, which returns the smallest integer not less than $x$. $\nabla {\vec e}$ denotes gradient of a vector ${\vec e}$. $\min(.)$
gives the minimum of the quantities passed as its argument. $\Re{(c)} \textrm{ and } \Im{(c)}$ denote the real and imaginary parts
of a complex number $c$, respectively. $\mathrm{Pr}(E)$ denotes the probability of event $E$. Any new or unconventional notation used in the paper is defined in the place where it occurs.
\section{System Setup}\label{sysetup}
\begin{figure}
\centering\includegraphics[width=0.8\columnwidth]{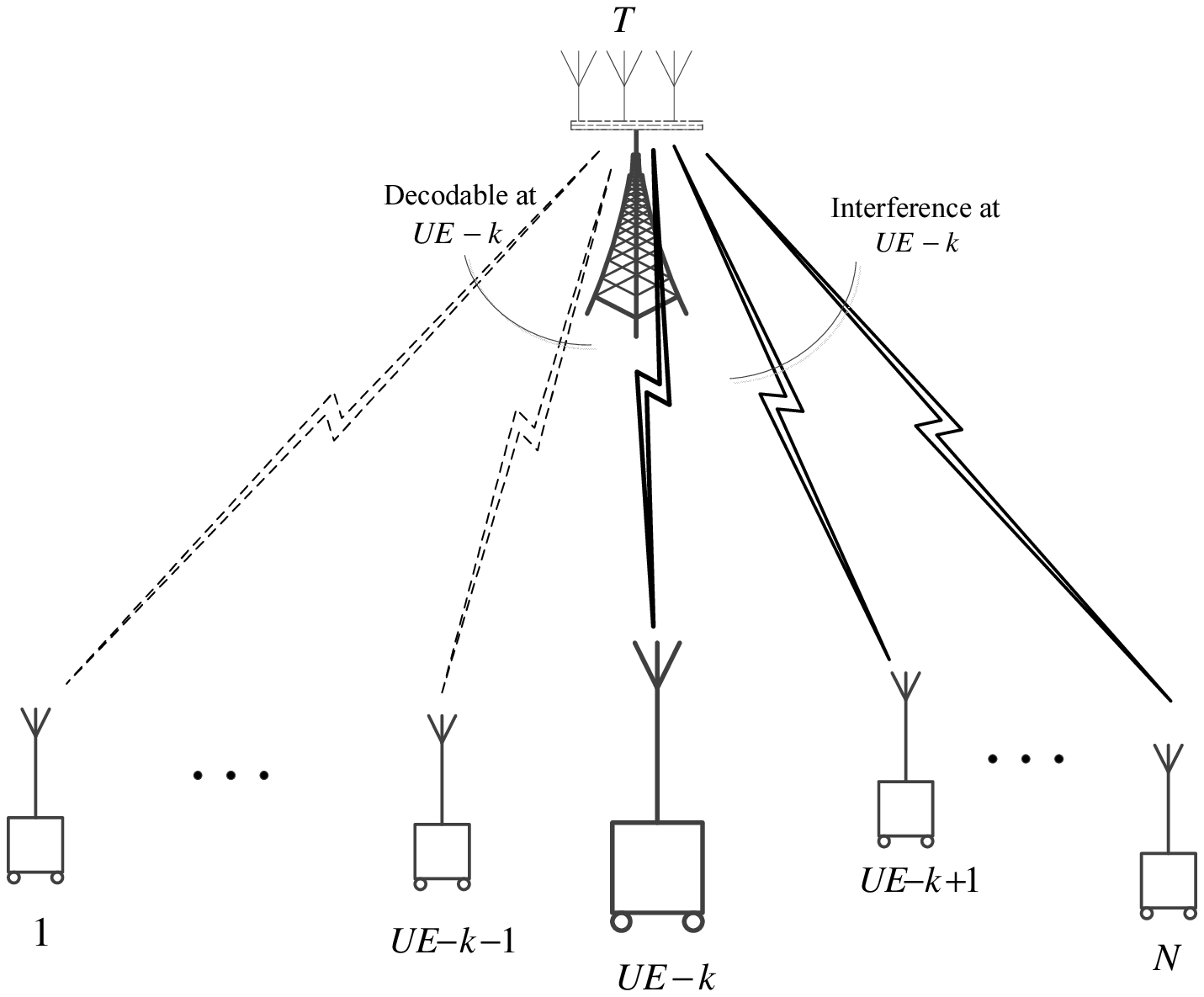}\protect\caption{The system setup. A BS with $T$ antennas serves $N$ users. The user UE-$k$ receives interference
from the users UE-$k+1$ to UE-$N$. The signals of remaining users from UE-$1$ to UE-$k-1$ are cancelled at UE-$k$.}
\label{fig:draw}
\end{figure}
We consider the downlink of a BS, equipped with $T$ antennas and serving $N$ single antenna users. NOMA principle is used for transmission purposes (please
refer to Fig. \ref{fig:draw}). We further assume that
the transmitted signal of each user equipment (UE) is linearly weighted with a complex vector. Specifically, to all $N$ users, the
BS transmits a superposition of the individual messages, ${\vec w}_is_i$ for all $i$, where ${\vec w}_i\in\mathbb{C}^{T}$ and $s_i$
are the complex weight vector and the transmitted symbol for UE-$i$, respectively. Therefore, under frequency flat channel conditions the
received signal $y_i$ at UE-$i$ is
\begin{align}
y_i&={\vec h}_i\herm\left(\sum_{j=1}^N{\vec w}_js_j\right)+n_i=\sum_{j=1}^N{\vec h}_i\herm{\vec w}_js_j+n_i,\quad i=1,\ldots,N,\label{sys2}
\end{align}
where ${\vec h}_i=\sqrt{d_i^{-\gamma}}{\vec g}_i\in\mathbb{C}^T$, with $d_i$ being the distance between $i^{\textrm{th}}$ UE and the BS, $\gamma$ is the path loss exponent,
${\vec g}_i\sim \mathcal{CN}(0,{\vec I})$, and $n_i$ represents circularly symmetric complex Gaussian noise with variance $\sigma^2$.
Subsequently, NOMA proposes to employ SIC at individual UEs, based on the particular ordering. For instance, the works in \cite{Saito,Ding1}
use the fact that for a single-input single-output (SISO) system, once the channels are arranged in a particular order (increasing or decreasing), then a UE-$k$ decodes all those
UE-$i$ signals, whose index $i<k$ (increasing order) and $i>k$ (decreasing order). An illustration of this process is also given in Fig. \ref{fig:draw}. However, simple SISO ordering
cannot be transformed to the MISO setup. The present work does not focus on the optimal ordering problem, but in the design of the complex weighting vectors, ${\vec w}_i$, that maximize
the aggregate throughput of the system, for a given UE ordering. Next, we assume that the channel state information (CSI) is perfectly known at all nodes.
\subsection{Problem Formulation}
We assume that the UE-$1$ is the weakest (and hence cannot decode any interfering signals), while UE-$N$ is the strongest user,
and is able to nullify all other UE interference by performing SIC. The other UEs are placed in an increasing order with respect to their
index numbers. For instance, UE-$m$ is placed before UE-$n$ if index $m<n$. Increasing
channel strengths can be used to order the users. But, as mentioned above, this ordering may not be optimal, and better rates may be achievable for different
order of users. According to
NOMA the achievable rate after SIC operation at the $k^{\textrm{th}}$ user, with $k>i$ for all $i=1,\ldots,k-1$, is \cite{Ding1,Saito}
\begin{align}
R_k^k=\log_2\left(1+\frac{|{\vec h}_k\herm{\vec w}_k|^2}{\sum_{j=k+1}^N|{\vec h}_k\herm{\vec w}_j|^2+\sigma^2}\right),\quad 1\leq k\leq N-1\label{rtk}.
\end{align}
An important observation should be noted here. For the above rate to be achievable at UE-$k$, it
is necessary for all UE-$j$, with $j>k$, to satisfy
\begin{align}
R_j^k=\log_2\left(1+\frac{|{\vec h}_j\herm{\vec w}_k|^2}{\sum_{m=k+1}^N|{\vec h}_j\herm{\vec w}_m|^2+\sigma^2}\right)\geq R_{th}\quad j=k+1,\ldots,N\label{rtk-cd1}
\end{align}
where $R_j^k$ is the rate of UE-$j$ to decode the message of $k^{\textrm{th}}$ UE, and $R_{th}$ is some target data rate for user $R_k^k$.
In addition, to allocate non-trivial data rates to the weaker users, which present a lower decoding capability in a given order, the following condition must also be satisfied
\begin{align}
|{\vec h}_k{\vec w}_1|^2\geq\ldots|{\vec h}_k{\vec w}_{k-1}|^2\geq|{\vec h}_k{\vec w}_k|^2\geq |{\vec h}_k{\vec w}_{k+1}|^2\ldots \geq|{\vec h}_k{\vec w}_N|^2.\label{rtk-cd}
\end{align}
\par As a further insight, \eqref{rtk-cd1} ensures that the signal-to-interference-plus-noise ratio (SINR) of UE-$j$ to decode the message of UE-$k$, where $j>k$, is higher compared to the SINR
of UE-$k$ to decode its own message. Once this condition is satisfied, all users, which are assumed to be
at a `higher' level in the given ordering, are able to perform SIC. Therefore, we propose to maximize the minimum
of these `direct' and `cross-user' decoding SINRs. To further exemplify, consider a three user system with UE-$1$ the lowest and the UE-$2$
the highest in the ordering. Now, assume that $\textrm{SINR}_1^1\geq T_{th}\textrm{ and }\textrm{SINR}_w^1<T_{th},\:w=2,3$, where $T_{th}$ is some threshold rate. In this scenario, both
users $2$ and $3$ are unable to decode the message of UE-$1$ as the $\textrm{SINR}_1^1$ is at least as large as
$T_{th}$, and therefore, SIC cannot be applied. Motivated by this, we aim at obtaining such precoders that ensure, we have
$T_{th}\leq \textrm{SINR}_w^1,\:w=1,2,3$. 
Moreover, the sequence of inequalities in \eqref{rtk-cd} helps to boost up the desired signal level of the `lower' level users, and in the absence of this guarantee it is likely that
most, if not all, radio resources are allocated to the users that receive very low or no interference.
The sum rate, $R_{sum}$, therefore, is given by
\begin{align}
R_{sum}=\sum_{k=1}^{N-1}\log_2\left(1+\min\left(\textrm{SINR}_k^k,\ldots,\textrm{SINR}_N^k\right)\right)+\log_2\left(1+\frac{|{\vec h}_N\herm{\vec w}_N|^2}{\sigma^2}\right),\label{Rsum}
\end{align}
where
\begin{align}
\textrm{SINR}_i^k=\frac{|{\vec h}_i\herm{\vec w}_k|^2}{\sum_{m=k+1}^N|{\vec h}_i\herm{\vec w}_m|^2+\sigma^2},\qquad i=1,\ldots,N.\label{sinrNK}
\end{align}
%
%
%
Now, the optimization problem can be formulated as
\begin{subequations}\label{Rprob0}
\begin{align}
&\underset{{\vec w}_i\in\mathbb{C}^{T},\forall i}{\operatorname{maximize}}& & R_{sum}\label{Rprob1}\\
&\operatorname{s.\;t.}& &
\hspace{-8mm}|{\vec h}_k{\vec w}_1|^2\geq\ldots|{\vec h}_k{\vec w}_{k-1}|^2\geq|{\vec h}_k{\vec w}_k|^2\geq |{\vec h}_k{\vec w}_{k+1}|^2\ldots \geq|{\vec h}_k{\vec w}_N|^2
,\quad 1\leq k\leq N\label{Rprob2}
\\&&&\sum_{i=1}^N\|{\vec w}_i\|_2^2\leq P_{th},\label{Rprob3}
\end{align}
\end{subequations}
where the constraint in \eqref{Rprob3} represents that the total power, which is upper bounded to $P_{th}$. It is important to mention here that in the original NOMA \cite{Saito,DOCOMOnoma},
its concept was applied only to two users, but it can be
extended to a multiuser setting. Such an extension requires optimal grouping \cite{Ding3} and hence it is left open for future investigation.
\section{Prerequisites}\label{prereq}
In order to solve the optimization problem in \eqref{Rprob0}, we will eventually present an iterative algorithm. However, first it is
necessary to transform the original problem and then to
apply approximations that render it tractability.
\subsection{Equivalent Transformations}
The problem in \eqref{Rprob0} is non-convex, and it seems that it is not possible to directly approximate it, since the only convex constraint is the power constraint. Therefore, several
steps need to be invoked before we can present an algorithm, which solves this problem approximately. To this end, we first introduce the
vector ${\vec r}\in\mathbb{R}^N_+$ and we observe that
\eqref{Rprob0} can be equivalently written as
\begin{subequations}\label{Rproba0}
\begin{align}
&\underset{{\vec w}_i\in\mathbb{C}^{T},\forall i,{\vec r}\in\mathbb{R}^N_+}{\operatorname{maximize}}& & \left(\prod_{k=1}^N r_k\right)^{\frac{1}{N}}\label{Rproba1}\\
&\operatorname{s. t.}& &
r_k-1\leq \min\left(\textrm{SINR}_k^k,\ldots,\textrm{SINR}_N^k\right),\quad k=1,\ldots,N-1\label{Rproba2}\\&&&
r_N-1\leq \frac{|{\vec h}_N\herm{\vec w}_N|^2}{\sigma^2}\label{Rproba3}\\&&&\textrm{\eqref{Rprob2}}\textrm{ \& \eqref{Rprob3}},\label{Rproba4}
\end{align}
\end{subequations}
where $r_i,i=1,\ldots,N$ are the components of ${\vec r}$, and the objective has been obtained by considering that $\log(\cdot)$ is a non-decreasing function, and the geometric mean
of the vector ${\vec r}$, i.e., $\left(\prod_{k=1}^N r_k\right)^{1/N}$, is concave and increasing\footnote{It is not necessary to explicitly constrain the vector ${\vec r}$
to be positive, since for non-zero data rates this condition holds.}. It is well known that the geometric mean can be readily expressible as a system of second-order cone (SOC)
constraints \cite{lobo}. So this step has no negative impact on the tractability of the objective function.  However, the overall problem still remains intractable. Nonetheless,
the original formulation is \emph{factored} into several different constraints, and so, these
factors can be processed individually. We first focus on the constraints in \eqref{Rproba2}, and then move to the remaining intractable constraints. Without loss of generality, it holds that
\begin{align}
r_k-1\leq \min\left(\textrm{SINR}_k^k,\ldots,\textrm{SINR}_N^k\right)
\Leftrightarrow r_k-1\leq \begin{cases}\textrm{SINR}_k^k\\\min\left(\textrm{SINR}_{k+1}^k,\ldots,\textrm{SINR}_N^k\right),\end{cases}\label{Rproba2eq0}
\end{align}
for $k=1,\ldots,N-1$. The constraint in \eqref{Rproba2} has been purposely written as that in \eqref{Rproba2eq0}, since the first term $\textrm{SINR}_k^k$ is different from the
remaining ones. Hence, it is necessary to deal with the first term and the remaining $N-k$ terms passed as argument of the $\min(\cdot)$ function.\par By introducing,
$\bar{{\vec w}}\in\mathbb{R}^{N-1}_+$, it holds that
\begin{align}
r_k-1\leq \frac{|{\vec h}_k\herm{\vec w}_k|^2}{\sum_{j=k+1}^N|{\vec h}_k\herm{\vec w}_j|^2+\sigma^2}
\Leftrightarrow\begin{cases}\bar{w}_kr_k-\bar{w}_k\leq |{\vec h}_k\herm{\vec w}_k|^2\\\sum_{j=k+1}^N|{\vec h}_k\herm{\vec w}_j|^2+\sigma^2\leq \bar{w}_k,\end{cases}\label{Rproba2eq01}
\end{align}
where $\bar{w}_k$ is the $k^{\textrm{th}}$ component of the vector $\bar{{\vec w}}$, and the expression of $\textrm{SINR}_k^k$ is used. Likewise, for an arbitrary
$\textrm{SINR}_j^k$, $k+1\leq j\leq N$, belonging to the remaining terms in the $\min(\cdot)$ function,
we introduce the new variable, ${\vec v}\in\mathbb{R}^{0.5(N^2-N)}_+$, and write the corresponding constraint as the following system of inequalities
\begin{align}
r_kv_j-v_j\leq |{\vec h}_j\herm{\vec w}_k|^2,\quad\sum_{m=k+1}^N|{\vec h}_j\herm{\vec w}_m|^2+\sigma^2\leq v_j,\label{Rproba2eq02}
\end{align}
where $v_j$ is the $j^{\textrm{th}}$ element of ${\vec v}$. Note, that even if the constraints in \eqref{Rproba2} have been transformed, the problem
remains intractable.\par From the inequalities in \eqref{Rprob2}, it holds that
\begin{align}
&|{\vec h}_k{\vec w}_1|^2\geq\ldots|{\vec h}_k{\vec w}_{k-1}|^2\geq|{\vec h}_k{\vec w}_k|^2\geq |{\vec h}_k{\vec w}_{k+1}|^2\ldots \geq|{\vec h}_k{\vec w}_N|^2\\&\Leftrightarrow
\begin{cases}|{\vec h}_k{\vec w}_N|^2\leq \min_{m\in[1,N-1]}|{\vec h}_k{\vec w}_{m}|^2\\\cdots\\|{\vec h}_k{\vec w}_{k+1}|^2\leq \min_{m\in[1,k]}|{\vec h}_k{\vec w}_{m}|^2\\ \cdots \\
|{\vec h}_k{\vec w}_{2}|^2\leq |{\vec h}_k{\vec w}_{1}|^2.
\end{cases}\triangleq \mathcal{T}(k,N)\label{Rprob2eq}
\end{align}
Similarly, equivalent transformations, $\mathcal{T}(1,N)$ and $\mathcal{T}(N,N)$ for $k=1,N$ can be obtained. We conclude this subsection by presenting the equivalent formulation of \eqref{Rprob0} as
\begin{subequations}\label{RprobaTR0}
\begin{align}
&\underset{\substack{{\vec w}_i\in\mathbb{C}^{T},\forall i,{\vec r}\in\mathbb{R}^N_+,\bar{{\vec w}}\in\mathbb{R}^{N-1}_+,\\{\vec v}\in\mathbb{R}^{0.5(N^2-N)}_+}}{\operatorname{maximize}}& &
\left(\prod_{k=1}^N r_k\right)^{\frac{1}{N}}\label{RprobaTR1}\\
&\operatorname{s.\;t.}& &
\begin{cases}\bar{w}_kr_k-\bar{w}_k\leq |{\vec h}_k\herm{\vec w}_k|^2\\\sum_{j=k+1}^N|{\vec h}_k\herm{\vec w}_j|^2+\sigma^2\leq \bar{w}_k\end{cases}\quad k=1,\ldots,N-1\label{RprobaTR2}\\&&&
r_kv_j-v_j\leq |{\vec h}_j\herm{\vec w}_k|^2,\quad\sum_{m=k+1}^N|{\vec h}_j\herm{\vec w}_m|^2+\sigma^2\leq v_j\quad j=k+1,\ldots,N\label{RprobaTR3}\\&&&
r_N-1\leq \frac{|{\vec h}_N\herm{\vec w}_N|^2}{\sigma^2}\label{RprobaTR4}\\&&& \mathcal{T}(1,N),\ldots,\mathcal{T}(k,N),\ldots,\mathcal{T}(N,N)\textrm{ \& \eqref{Rprob3}}\label{RprobaTR5}.
\end{align}
\end{subequations}
\subsection{Approximation of the non-convex constraints}
Next, we approximate the equivalent formulation in \eqref{RprobaTR0}. To this end, note that, excluding the power constraint,
the first set of constraints in \eqref{RprobaTR2}, \eqref{RprobaTR3}, and the constraints in \eqref{RprobaTR4} and \eqref{RprobaTR5} are all non-convex.
The rest of the constraints are convex, and in fact admit SOC representation. Consider the second set of constraints in \eqref{RprobaTR2} i.e.,
\begin{align}
\sum_{j=k+1}^N|{\vec h}_k\herm{\vec w}_j|^2+\sigma^2\leq \bar{w}_k\Leftrightarrow
\left\Vert\begin{pmatrix}{\vec h}_k\herm{\vec w}_{k+1} \\ \vdots \\ {\vec h}_k\herm{\vec w}_{N} \\ \sigma \\ \frac{\bar{w}_k-1}{2}\end{pmatrix}\right\Vert_2\leq \frac{\bar{w}_k+1}{2},\quad k=1,\ldots,N-1.\label{RprobaTR2sec}
\end{align}
Similarly,
\begin{align}
\sum_{m=k+1}^N|{\vec h}_j\herm{\vec w}_m|^2+\sigma^2\leq v_j\Leftrightarrow
\left\Vert\begin{pmatrix}{\vec h}_j\herm{\vec w}_{k+1} \\ \vdots \\ {\vec h}_j\herm{\vec w}_{N} \\ \sigma \\ \frac{v_j-1}{2}\end{pmatrix}\right\Vert_2\leq \frac{v_j+1}{2},
\quad j=k+1,\ldots,N.\label{RprobaTR3sec}
\end{align}
Now, in order to tackle the non-convex constraints, the so called CCP is used. The CCP has been widely used in neural computing \cite{CCPTut}, and has recently
found applications in wireless signal processing \cite{HanifTSP,TranSPL}. The CCP has also been referred to as minorization-maximization algorithm (MMA) \cite{MMATut,MMA2}.\par
First, the procedure of handling the first set of non-convex constraints in \eqref{RprobaTR2} is considered. The approximation of the other non-convex constraints closely follows the
same technique. Consider the $k^{\textrm{th}}$ constraint
\begin{align}
\bar{w}_kr_k-\bar{w}_k\leq |{\vec h}_k\herm{\vec w}_k|^2.\label{RprobaTR2NCvx1}
\end{align}
This is non-convex because of the bilinear term on the left side and the quadratic term on the right side of the
inequality. An equivalent transformation of the above inequality is
\begin{align}
\bar{w}_kr_k-\bar{w}_k\leq (\theta_{k,k}^i)^2+(\theta_{k,k}^r)^2=\|\boldsymbol{\uptheta}_{k,k}\|_2^2,
\quad \theta_{k,k}^r=\Re\left({\vec h}_k\herm{\vec w}_k\right),\theta_{k,k}^i=\Im\left({\vec h}_k\herm{\vec w}_k\right)\label{RprobaTR2NCvx2}
\end{align}
where $\boldsymbol{\uptheta}_{k,k}=[\theta_{k,k}^r,\theta_{k,k}^i]\trans$ and $f(\boldsymbol{\uptheta}_{k,k})\triangleq |{\vec h}_k\herm{\vec w}_k|^2$. Since the function in
the right side of \eqref{RprobaTR2NCvx2} is a convex one, it follows that \cite{boyd}
\begin{align}
f(\boldsymbol{\uptheta}_{k,k})=\|\boldsymbol{\uptheta}_{k,k}\|_2^2\geq \|\boldsymbol{\uptheta}_{k,k}^t\|_2^2+2(\boldsymbol{\uptheta}_{k,k}^t)\trans
(\boldsymbol{\uptheta}_{k,k}-\boldsymbol{\uptheta}_{k,k}^t)\triangleq g(\boldsymbol{\uptheta}_{k,k},\boldsymbol{\uptheta}_{k,k}^t),\label{RprobaTR2NCvx3}
\end{align}
where the right side of the inequality in \eqref{RprobaTR2NCvx3} is the first order Taylor approximation of the function $\|\boldsymbol{\uptheta}_{k,k}\|_2^2$ around
$\boldsymbol{\uptheta}_{k,k}^t$. Clearly, this formulation is linear in the variable $\boldsymbol{\uptheta}_{k,k}$, and will be used instead of the
original norm-squared function. Three important properties follow here
\begin{subequations}\label{propgTh0}
\begin{align}
&f(\boldsymbol{\uptheta}_{k,k})\geq g(\boldsymbol{\uptheta}_{k,k},\boldsymbol{\uptheta}_{k,k}^t), \quad \textrm{ for all } \boldsymbol{\uptheta}_k\label{propgTh1}\\&
f(\boldsymbol{\uptheta}_{k,k}^t)=g(\boldsymbol{\uptheta}_{k,k}^t,\boldsymbol{\uptheta}_{k,k}^t),\label{propgTh2}\\&
\nabla f(\boldsymbol{\uptheta}_{k,k})|_{\boldsymbol{\uptheta}_{k,k}^t}=\nabla g(\boldsymbol{\uptheta}_{k,k},\boldsymbol{\uptheta}_{k,k}^t)|_{\boldsymbol{\uptheta}_{k,k}^t}\label{propgTh3}
\end{align}
\end{subequations}
where the notation $(\cdot)|_{_{\boldsymbol{\uptheta}_{k,k}^t}}$ is used to represent the value of the function at ${\boldsymbol{\uptheta}_{k,k}^t}$. The basic idea of the
approximation algorithm presented below is to maximize the
minorant $g(\boldsymbol{\uptheta}_{k,k},\boldsymbol{\uptheta}_{k,k}^t)$ over the variable $\boldsymbol{\uptheta}_{k,k}$, in order
to obtain the next iterate term, $\boldsymbol{\uptheta}_{k,k}^{t+1}$, i.e.,
\begin{align}
\boldsymbol{\uptheta}_{k,k}^{t+1}=\max_{\boldsymbol{\uptheta}_{k,k}}g(\boldsymbol{\uptheta}_{k,k},\boldsymbol{\uptheta}_{k,k}^t)\label{updtTh0}.
\end{align}
Using these considerations, it can be easily concluded that
\begin{align}
f(\boldsymbol{\uptheta}_{k,k}^{t+1})&= f(\boldsymbol{\uptheta}_{k,k}^{t+1})-g(\boldsymbol{\uptheta}_{k,k}^t,\boldsymbol{\uptheta}_{k,k}^{t+1})
+g(\boldsymbol{\uptheta}_{k,k}^t,\boldsymbol{\uptheta}_{k,k}^{t+1})\label{serInq0}\\&\stackrel{(a)}{\geq} g(\boldsymbol{\uptheta}_{k,k}^t,\boldsymbol{\uptheta}_{k,k}^{t+1})
\stackrel{(b)}{\geq} g(\boldsymbol{\uptheta}_{k,k}^t,\boldsymbol{\uptheta}_{k,k}^{t})\stackrel{(c)}{=}f(\boldsymbol{\uptheta}_{k,k}^{t}),\label{serInq1}
\end{align}
where $(a)$ follows from $f(\boldsymbol{\uptheta}_{k,k})\geq g(\boldsymbol{\uptheta}_{k,k},\boldsymbol{\uptheta}_{k,k}^t)$, $(b)$ is due to \eqref{updtTh0}, and
the final equality $(c)$ is due to \eqref{propgTh2}.\par
Now, to deal with the
bilinear product on the left side of \eqref{RprobaTR2NCvx1}, first we observe that for nonnegative  $\bar{w}_k,r_k$ it holds that
\begin{align}
\bar{w}_kr_k=\frac{1}{4}\left[(\bar{w}_k+r_k)^2-(\bar{w}_k-r_k)^2\right].\label{RprobaTR2NCvx4}
\end{align}
The quadratic term being subtracted in the above inequality can be well approximated by a first order Taylor series
around $\bar{w}_k^t,r_k^t$. Thus, the overall constraint in \eqref{RprobaTR2NCvx1} reads as
\begin{align}
0.25(\bar{w}_k+r_k)^2-\bar{w}_k-0.25\left[(\bar{w}_k^t-r_k^t)^2+2(\bar{w}^t_k-r^t_k)\{\bar{w}_k-\bar{w}^t_k-
r_k+r_k^t\}\right]\leq g(\boldsymbol{\uptheta}_{k,k},\boldsymbol{\uptheta}_{k,k}^t),\label{RprobaTR2NCvx5}
\end{align}
which is convex in the variables of interest.\par Following similar procedure the
remaining non-convex constraints in \eqref{RprobaTR3}, \eqref{RprobaTR4} and \eqref{RprobaTR5} can be approximated as follows. The $j$-th constraint in
\eqref{RprobaTR3} and that in \eqref{RprobaTR4} can be written as
\begin{align}
&0.25(r_k+v_j)^2-v_j-0.25\left[(r_k^t-v_j^t)^2+2(r^t_k-v^t_j)\{r_k-r^t_k-
v_j+v_j^t\}\right]\label{RprobaTR3SG}\\&\leq \bar{g}(\boldsymbol{\uptheta}_{j,k},\boldsymbol{\uptheta}_{j,k}^t)
\sigma^2(r_N-1)\leq \bar{g}(\boldsymbol{\uptheta}_{N,N},\boldsymbol{\uptheta}_{N,N}^t),\label{RprobaTR4SG}
\end{align}
where
\begin{align}
&\boldsymbol{\uptheta}_{j,k}=[\theta_{j,k}^r,\theta_{j,k}^i]\trans,\boldsymbol{\uptheta}_{N,N}=[\theta_{N,N}^r,\theta_{N,N}^i]\trans,
\bar{g}(\boldsymbol{\uptheta}_{j,k},\boldsymbol{\uptheta}_{j,k}^t)=
\|\boldsymbol{\uptheta}_{j,k}^t\|_2^2+2(\boldsymbol{\uptheta}_{j,k}^t)\trans
(\boldsymbol{\uptheta}_{j,k}-\boldsymbol{\uptheta}_{j,k}^t),\notag\\&\bar{g}(\boldsymbol{\uptheta}_{N,N},\boldsymbol{\uptheta}_{N,N}^t)=
\|\boldsymbol{\uptheta}_{N,N}^t\|_2^2+2(\boldsymbol{\uptheta}_{N,N}^t)\trans
(\boldsymbol{\uptheta}_{N,N}-\boldsymbol{\uptheta}_{N,N}^t)\notag
\end{align} and $r_k^t,v_j^t,\boldsymbol{\uptheta}_{j,k}^t,\boldsymbol{\uptheta}_{N,N}^t$
represent the points around which the quadratic terms have been linearized.\par Finally, the last set of non-convex constraints in \eqref{RprobaTR5} can
be tackled similarly. To demonstrate it, we linearize the first set of constraints in \eqref{Rprob2eq}, i.e.,
\begin{align}
|{\vec h}_k{\vec w}_N|^2\leq \min_{m\in[1,N-1]}\tilde{g}(\boldsymbol{\upphi}_{k,m},\boldsymbol{\upphi}_{k,m}^t),
\end{align}
where
\begin{align}
\boldsymbol{\upphi}_{k,m}=[\phi_{k,m}^r,\phi_{k,m}^i]\trans, \tilde{g}(\boldsymbol{\upphi}_{k,m},\boldsymbol{\upphi}_{k,m}^t)=
\|\boldsymbol{\upphi}_{k,m}^t\|_2^2+2(\boldsymbol{\upphi}_{k,m}^t)\trans
(\boldsymbol{\upphi}_{k,m}-\boldsymbol{\upphi}_{k,m}^t)\notag
\end{align}
and $\boldsymbol{\upphi}_{k,m}^t$ is the linearization point. The notation $\bar{\mathcal{T}}(k^t,N^t)$ is used to represent the approximation of
the remaining inequalities using this procedure.
\section{The Proposed Solution}\label{proSol}
Having set up the stage as above, in this section the procedure that
provides a tractable approximation to the sum rate maximization problem is outlined.
\subsection{The Procedure}
Using the above equivalent transformations and approximations, in the $t^{\textrm{th}}$ iteration of the algorithm
outlined in Table \ref{ALG}, the following optimization problem is solved
\begin{subequations}
\label{TthRun0}
\begin{align}
&\underset{\substack{{\vec w}_i\in\mathbb{C}^{T},\forall i,{\vec r}\in\mathbb{R}^N_+,\bar{{\vec w}}\in\mathbb{R}^{N-1}_+,\\{\vec v}\in\mathbb{R}^{0.5(N^2-N)}_+,\hspace{1 mm}\mathcal{A}}}
{\operatorname{maximize}}& & \left(\prod_{k=1}^N r_k\right)^{\frac{1}{N}}\label{TthRun1}\\
(Pb_t)\hspace{10 mm}&\operatorname{s.t.}& &
\textrm{\eqref{RprobaTR2sec} \& \eqref{RprobaTR2NCvx5}},\quad k=1,\ldots,N-1\label{TthRun2}\\&&&
\textrm{\eqref{RprobaTR3sec} \& \eqref{RprobaTR3SG}}\quad j=k+1,\ldots,N,\quad
\textrm{\eqref{RprobaTR4SG}}\label{TthRun3}\\&&& \bar{\mathcal{T}}(1^t,N^t),\ldots,\bar{\mathcal{T}}(k^t,N^t),\ldots,\bar{\mathcal{T}}(N^t,N^t)\textrm{ \& \eqref{Rprob3}},\label{TthRun4}
\end{align}
\end{subequations}
where for all $j,k,m,\mathcal{A}\triangleq\{\boldsymbol{\uptheta}_{k,k}\in\mathbb{R}^{2N-2},\boldsymbol{\uptheta}_{j,k}\in\mathbb{R}^{N^2-N},
\boldsymbol{\uptheta}_{N,N}\in\mathbb{R}^2,\boldsymbol{\upphi}_{k,m}\in\mathbb{R}^{2N^2-2N}\}$ represents
the collection of all auxiliary variables. For the sake of notational convenience, all parameters about which the
quadratic terms are linearized in iterate $t$ are defined as
\begin{align}
\boldsymbol{\Uplambda}^t\triangleq [\bar{w}_k^t,r_k^t,r_k^t,v_j^t,\boldsymbol{\uptheta}_{k,k}^t,\boldsymbol{\uptheta}_{j,k}^t,\boldsymbol{\uptheta}_{N,N}^t,\boldsymbol{\upphi}_{k,m}^t]
\label{LumpVARt}.
\end{align}

The MMA (CCP) algorithm used to solve \eqref{TthRun0} has been summarized in Table \ref{ALG}. Note, that the convergence criteria can vary. For NOMA
sum rate maximization, this algorithm terminates when the difference between two successive values of sum rate is less than a
threshold. This aspect is discussed in more detail in section \ref{NumRes}.
\begin{table}[t]
\renewcommand{\arraystretch}{1.3}
\caption{NOMA/MISO Sum Rate Maximization} \centering
\begin{tabular}{ | l |}
\hline
\textbf{given} randomly generated $\boldsymbol{\Uplambda}^0$ feasible to \eqref{Rprob0}.\\
$t:=0$.\\
\textbf{repeat}\\
\hspace{5 mm}1- \emph{Solve} \eqref{TthRun0} labelled as $(Pb_t)$.\\
\hspace{5 mm}2- \emph{Set} $\boldsymbol{\Uplambda}^{t+1}=\boldsymbol{\Uplambda}^t$.\\
\hspace{5 mm}3- \emph{Update} $t:=t+1$.\\
\textbf{until} convergence or required number of iterations.\\\hline
\end{tabular}
\label{ALG}
\end{table}
\subsection{Properties of the Proposed Algorithm}
Before describing various characteristics of the algorithm presented above, let us define the feasible set, the objective
and the set of optimization variables in the $t^{\textrm{th}}$ iteration, respectively, as
\begin{align}
&\mathcal{F}_t=[{\vec w}_i \textrm{ for all } i,{\vec r},\bar{{\vec w}},{\vec v},\mathcal{A}| \textrm{constraints in } (Pb_t) \textrm{ are satsified} ]\label{feasSet}\\&
\mathcal{O}_t=\max\left[\sum_{k=1}^N r_k|\left\{{\vec w}_i \textrm{ for all } i,{\vec r},\bar{{\vec w}},{\vec v},\mathcal{A}\right\}\in\mathcal{F}_t\right]\label{Objt}\\&
\mathcal{V}_t=[{\vec w}_i \textrm{ for all } i,{\vec r},\bar{{\vec w}},{\vec v},\mathcal{A}].\label{Vart}
\end{align}
\subsubsection{Convergence}
\newtheorem{proposition}{Proposition}
\begin{proposition}\label{Feast}
The sequence of variables $\{\mathcal{V}_t\}_{t\geq 0}$ is feasible i.e., it belongs to $\mathcal{F}_0$, where $\mathcal{F}_0$
is the feasibility set of the original problem \eqref{RprobaTR0}.
\end{proposition}
\begin{IEEEproof}
Please, refer to Appendix \ref{proofFeast}.
\end{IEEEproof}
\begin{proposition}\label{Ascent}
The algorithm in Table \ref{ALG} returns a non-decreasing sequence of objective values i.e., $\mathcal{O}_{t+1}\geq\mathcal{O}_t$, and hence it converges.
\end{proposition}
\begin{IEEEproof}
Please, refer to Appendix \ref{proofAscent}.
\end{IEEEproof}
\par
In Proposition \ref{Ascent} the property \eqref{updtTh0} is used. It is important to note that the outcome of this proposition remains
valid as long as the surrogate is increasing and does not rely on explicit maximization. The increasing behaviour of all $SF$s can
be shown by following arguments similar to those outlined in \cite{HanifTSP,amir}. As a remark, we point out that when the feasibility set is
convex and compact, the algorithm converges to a finite value.
\subsubsection{KKT Conditions}
Under a couple of technical assumptions the accumulation point of the algorithm satisfies the KKT conditions, as summarized in the proposition given below.
\begin{proposition}\label{KKT-conv}
As the iteration number $t$ tends to infinity, the algorithm in Table \ref{ALG} converges to the KKT point of \eqref{RprobaTR0}.
\end{proposition}
\begin{IEEEproof}
Please, refer to Appendix \ref{proofKKT}.
\end{IEEEproof}
\section{A Reduced Complexity Approximation}\label{reDcomp}
In the original sum rate function given in \eqref{Rsum}, it has been ensured that users with high SNRs are able to
decode the messages of the weaker ones in the superposition coded signal, and hence apply SIC to remove interference from them.
Optimal ordering of users depends upon physical parameters like, transmit antennas, precoding vectors etc. However, in certain situations the channel ordering alone may
be sufficient to support the stronger users to decode the weaker ones.
In scenarios, this is true, only the first term in the $\min(\cdot)$ function for a
user $k$ needs to be retained and the objective becomes
\begin{align}
R^\prime_{sum}=\sum_{k=1}^{N-1}\log_2(1+\textrm{SINR}_k^k)+\log_2\left(1+\frac{|{\vec h}_N\herm{\vec w}_N|^2}{\sigma^2}\right).\label{ObjN}
\end{align}
From \eqref{ObjN} it can be seen that $N(N-1)/2$ SINR terms do not appear in the simplified sum rates. In turn, this means that
in the formulation of \eqref{RprobaTR0}, there are not $N^2-N$ inequality constraints, and clearly, a complexity improvement is expected.
Before moving on to the complexity analysis section, for completeness, the updated optimization problem solved in the $t^{\textrm{th}}$ run of the algorithm in Table \ref{ALG}, can be
written as
\begin{subequations}
\label{simpTthRun0}
\begin{align}
&\underset{\substack{{\vec w}_i\in\mathbb{C}^{T},\forall i,{\vec r}\in\mathbb{R}^N_+,\bar{{\vec w}}\in\mathbb{R}^{N-1}_+,\\\mathcal{A}_p}}
{\operatorname{maximize}}& & \left(\prod_{k=1}^N r_k\right)^{\frac{1}{N}}\label{simpTthRun1}\\
(Pb_t^\prime)\hspace{10 mm}&\operatorname{s.t.}& &
\textrm{\eqref{RprobaTR2sec} \& \eqref{RprobaTR2NCvx5}},\quad k=1,\ldots,N-1,\quad \textrm{\eqref{RprobaTR4SG}}\label{simpTthRun2}\\&&&
\bar{\mathcal{T}}(1^t,N^t),\ldots,\bar{\mathcal{T}}(k^t,N^t),\ldots,\bar{\mathcal{T}}(N^t,N^t)\textrm{ \& \eqref{Rprob3}},\label{simpTthRun4}
\end{align}
\end{subequations}
where now $\mathcal{A}_p\triangleq\{\boldsymbol{\uptheta}_{k,k},\boldsymbol{\uptheta}_{N,N},\boldsymbol{\upphi}_{k,m}\}$ has a reduced cardinality compared to the original
set of the variable set $\mathcal{A}$.\par In order to provide more insight into the approximation used above let us consider the following lemma.
\newtheorem{lemma}{Lemma}
\begin{lemma}\label{chnCond}
Suppose that
${\vec h}_{k+1}=c_{k+1}{\vec h}_k, k=1,\ldots,N-1$, so that ${\vec h}_{n}=c_{n}c_{n-1}\ldots c_{k+1}{\vec h}_k$, where $k+1\leq n\leq N$ and the
magnitudes of the complex constants $c_{n}c_{n-1}\ldots c_{k+1}\triangleq c_{n}^{k+1}$ is greater than one.
Under this assumption, when \eqref{Rsum} reduces to \eqref{ObjN}, then
\begin{align}\|{\vec h}_1\|_2<\|{\vec h}_2\|_2\ldots <\|{\vec h}_N\|_2.\end{align} 
\end{lemma}
\begin{IEEEproof}
Please, refer to Appendix \ref{proofchnCond}.
\end{IEEEproof}
\begin{figure}
\centering\includegraphics[width=1\columnwidth]{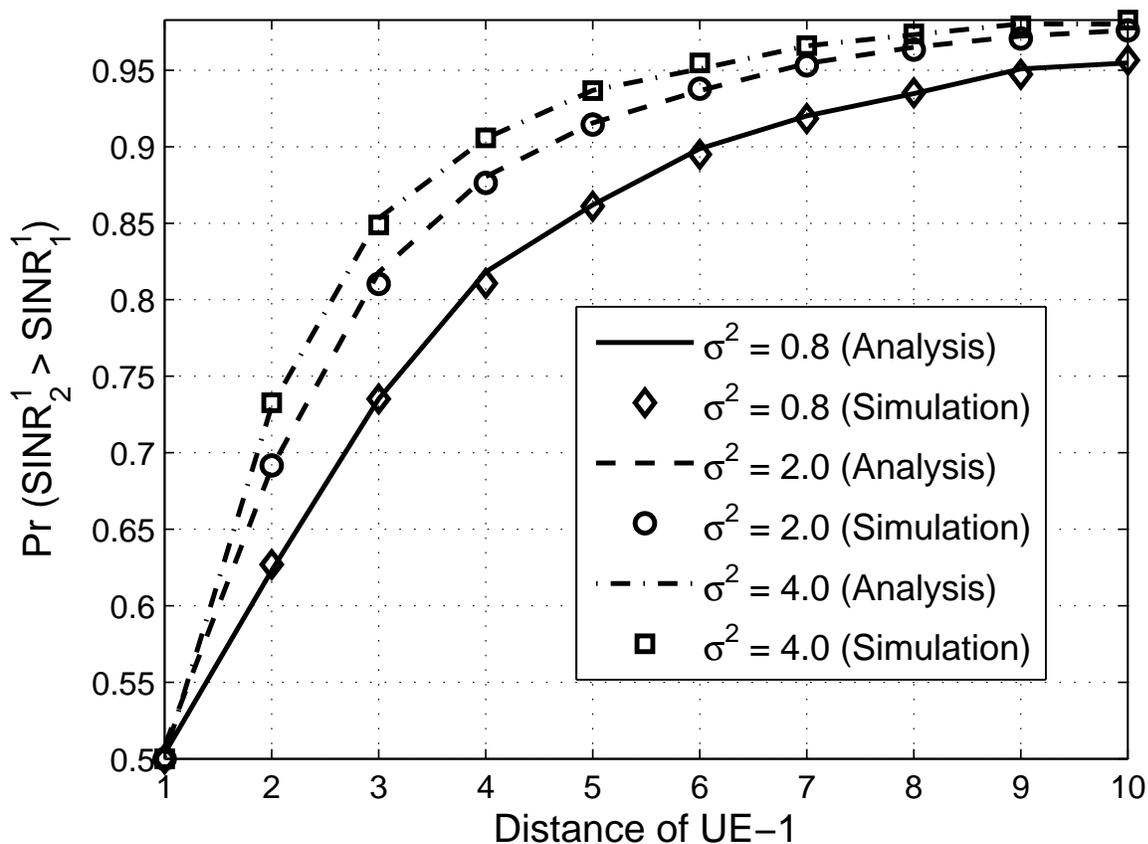}\protect\caption{Variation of $\mathrm{Pr}(\textrm{SINR}_2^1>\textrm{SINR}_1^1)$ with the distance of UE-1. $T=6,N=4,\gamma=2.0$ and
$\textrm{distance of UE-2}=1$ is fixed.}
\label{fig:probZguo}
\end{figure}
\par
From Lemma \ref{chnCond} it can be expected that, at least approximately, when that channels are clearly ordered, i.e., the magnitudes of successive
channel vectors differ significantly and the channel ratio inequalities as given above are satisfied, the problems in \eqref{simpTthRun0} and \eqref{TthRun0} are equivalent.
To further highlight, we evaluate below the probability of an event of interest.
\begin{lemma}\label{chnCondanother}
Consider a random unitary precoding matrix, i.e.,
$\mathbf{W}\herm\mathbf{W}=\mathbf{I}$,\footnote{Extending this lemma without the orthogonality constraint remains an open problem.} where ${\vec W}=[{\vec w}_1,\ldots,{\vec w}_N]$ and
$\mathbf{W}$ is independent of the channel matrices.
For $i\geq j$
\begin{align}
\mathrm{Pr}\left(\textrm{SINR}_i^k>\textrm{SINR}_j^k\right)
\end{align}
is given by
\begin{align}\label{probRequired}
\mathrm{Pr}\left(\textrm{SINR}_i^k>\textrm{SINR}_j^k\right)
&=1-e^{(\lambda_i+\lambda_j)\sigma^2} \lambda_i\sigma^2  \psi\left((\lambda_i+\lambda_j)\sigma^2,2(N-k)\right)\\
\notag&- e^{(\lambda_i+\lambda_j) \sigma^2}(N-k)
\psi\left((\lambda_i+\lambda_j) \sigma^2,2(N-k)+1\right),
\end{align}
where
\begin{align}
\psi(\lambda,m) = (-1)^{m} \frac{\lambda^{m-1}\textrm{Ei}(-\lambda)}{(m-1)!}+e^{-\lambda}\sum^{m-2}_{l=0}
\frac{(-1)^l\lambda^l}{(m-1)\cdots (m-1-l)}
\end{align}
and $\textrm{Ei}(x)$ is the exponential integral \cite{intg}.
\end{lemma}
\begin{IEEEproof}
Please, refer to Appendix \ref{proofchnCondanother}.
\end{IEEEproof}\par
When $\lambda_j>>\lambda_i$, \eqref{probRequired} can be approximated as
\begin{align}\label{zhi_probabAPP}
\mathrm{Pr}\left(\textrm{SINR}_i^k>\textrm{SINR}_j^k\right)
&\approx 1-e^{\lambda_j\sigma^2} \lambda_i\sigma^2  \psi\left(\lambda_j\sigma^2,2(N-k)\right)\\
\notag&- e^{\lambda_j \sigma^2}(N-k)
\psi\left(\lambda_j \sigma^2,2(N-k)+1\right).
\end{align}
\par It can be seen from \eqref{zhi_probabAPP} that when $d_i$ decreases (and thus $\lambda_i$ decreases), $\mathrm{Pr}\left(\textrm{SINR}_i^k>\textrm{SINR}_j^k\right)$ increases as well.
Hence, the stronger the channel of UE-$i$ compared to UE-$j$, the higher the probability $\mathrm{Pr}\left(\textrm{SINR}_i^k>\textrm{SINR}_j^k\right)$.
In order to further investigate the probability under consideration, in Fig. \ref{fig:probZguo}, the variation of $\mathrm{Pr}(\textrm{SINR}_2^1>\textrm{SINR}_1^1)$
is depicted in terms of the distance of UE-$1$. With a decrease in the channel strength of UE-1, $\mathrm{Pr}(\textrm{SINR}_2^1>\textrm{SINR}_1^1)$ increases, thereby justifying the use of $\textrm{SINR}_1^1$
instead of $\textrm{SINR}_2^1$. The probability $\mathrm{Pr}(\textrm{SINR}_2^1>\textrm{SINR}_1^1)$ varies inversely with the
noise variance for a given UE-1 distance. It can also be seen that for higher values of noise variance, the interference term dominates and the probability that
$\textrm{SINR}_1^1$ remains below $\textrm{SINR}_2^1$ is increased. In addition, this figure also validates the analytical results derived above.
%
%
%
\subsection{Complexity}
\begin{figure}
\centering\includegraphics[width=1\columnwidth]{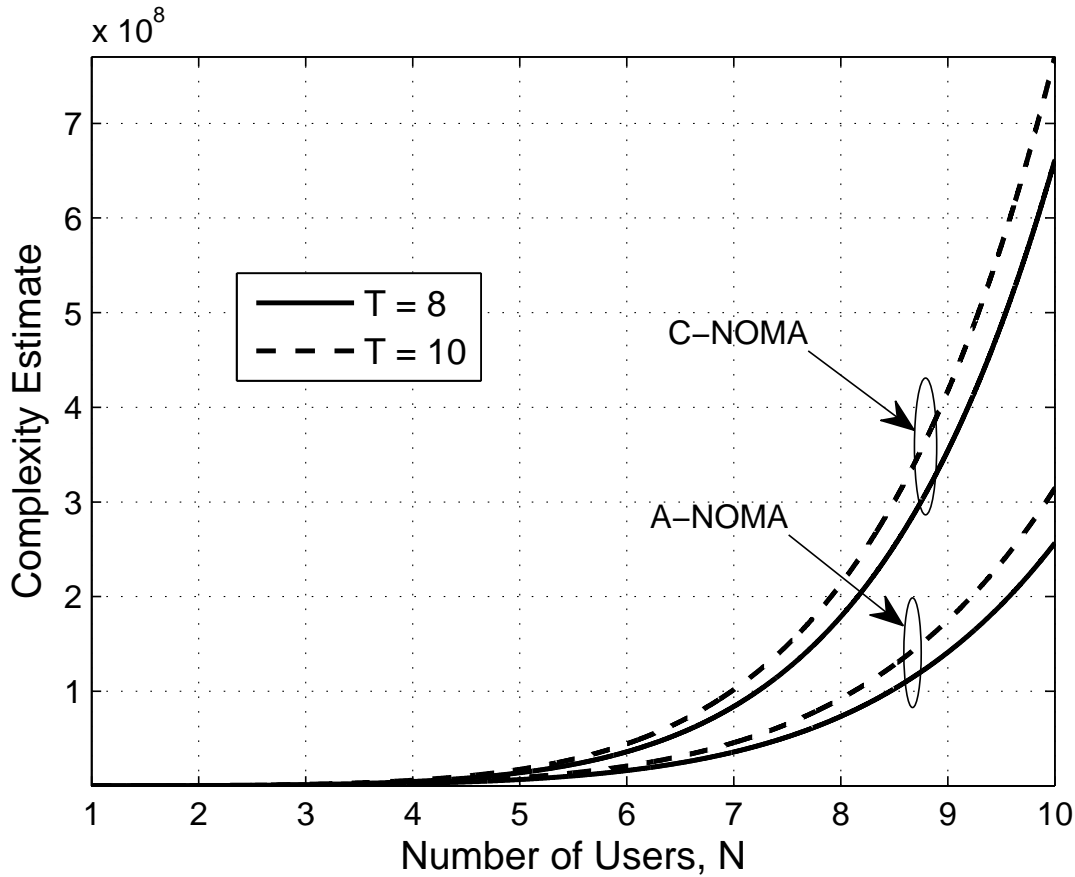}\protect\caption{Variation of the per iteration complexity of the
exact and approximate NOMA formulations with the number of users $N$ for given values of $T$ and $c=10$.}
\label{fig:complex}
\end{figure}
In each iteration of the procedure presented in Table \ref{ALG}, we solve an SOCP. The total number of iterations are fixed and only variables are updated in
each run of the algorithm. Hence, the worst case regarding the complexity is determined by the SOCP in each step. Therefore, to provide a complexity estimate,
the worst case complexity of the SOCP given by \eqref{TthRun0} or \eqref{simpTthRun0} is estimated. It is well known that for general interior-point methods
the complexity of the SOCP depends upon the number of constraints, variables and the dimension of each SOC constraint \cite{lobo}. The total number of constraints in the
formulations of \eqref{TthRun0} and \eqref{simpTthRun0} are $0.5N^3+0.5N^2+2N+c$ and $0.5N^3-0.5N^2+3N+c$, respectively, where
the non-negative integer constant, $c$, refers to the SOC constraints with different $N$. This happens because of the equivalent SOC representation of the
geometric mean, given in the objective function, also see \cite{lobo}. Therefore, for both problems the number of iterations needed to reduce the
duality gap to a small constant is upper bounded by $\mathcal{O}(\sqrt{0.5N^3+0.5N^2+2N+c})$ and $\mathcal{O}(\sqrt{0.5N^3-0.5N^2+3N+c})$, respectively \cite{lobo}. In order to calculate the
dimension of all SOCs in \eqref{TthRun0} we provide an upper bound because the sums of the dimensions for some constraints have been bounded from above by definite
integrals of increasing functions. This estimate is found to be $\lceil 1.833N^3+3N^2+8N+NT+3c-5.83333\rceil$ for \eqref{TthRun0}. The interior-point method's
per iteration worst case complexity estimate of \eqref{TthRun0} is $\mathcal{O}\left((3.5N^2+1.5N+2NT+c-1)^2 (\lceil 1.833N^3+3N^2+8N+NT+3c-5.83333\rceil)\right)$, where $3.5N^2+1.5N+2NT+c-1$ is the
number of optimization variables in \eqref{TthRun0}. Likewise, the interior-point method's per iteration complexity to solve the SOCP in \eqref{simpTthRun0} is given by
$\mathcal{O}((2N^2+3N+2NT+c-1)^2(1.5N^3-N^2+10.5N+NT+3c-4))$, where $2N^2+3N+2NT+c-1$ and $1.5N^3-N^2+10.5N+NT+3c-4$ are the optimization variables and the
total dimension of the SOC constraints in \eqref{simpTthRun0}.\par To provide further insight, we plot
the per iteration complexity estimates of the SOCPs in Fig. \ref{fig:complex}. The SOCP in \eqref{TthRun0} is called complete NOMA (C-NOMA), while the one in \eqref{simpTthRun0}
is dubbed as approximate NOMA (A-NOMA). The figure quantifies the increase in the complexity as a function of both $N$ and $T$.
\section{Numerical Results}\label{NumRes}
\begin{figure}
\centering\includegraphics[width=1\columnwidth]{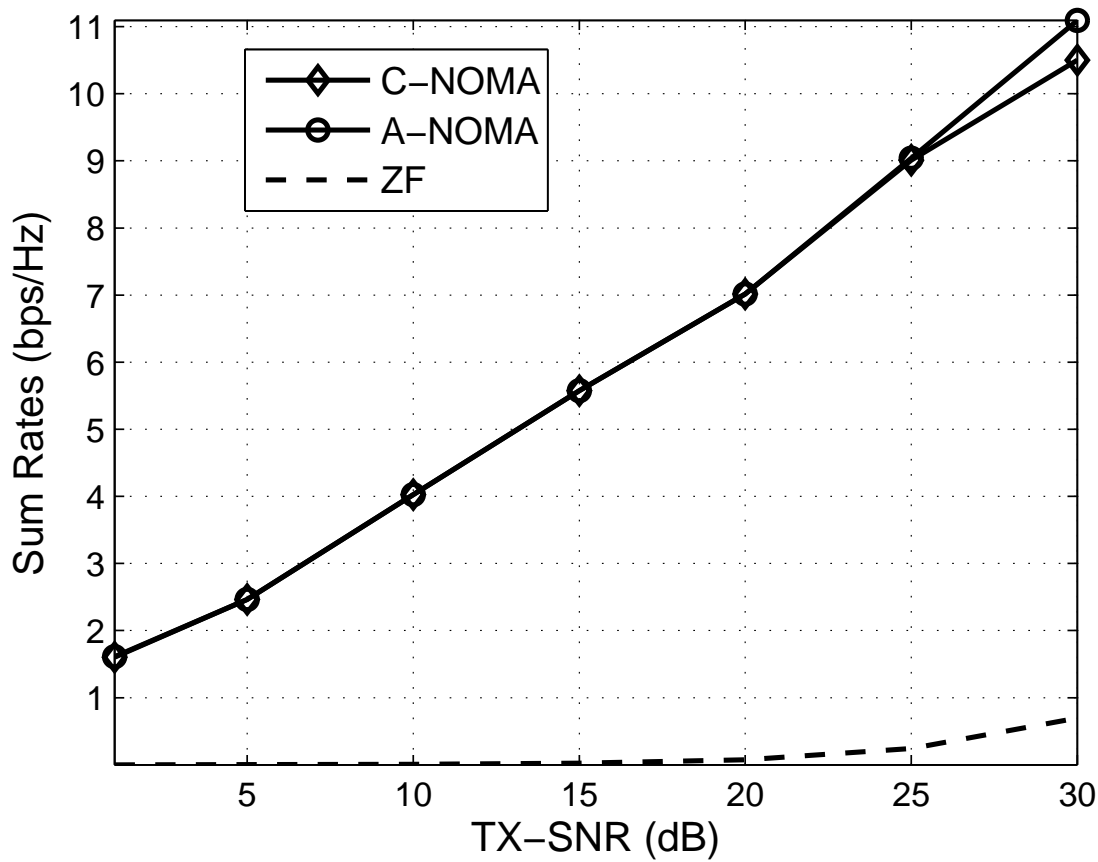}\protect\caption{Achievable sum rates vs. normalized transmit power called TX-SNR. We take $T=N=3, D_0=50,\gamma=2 \textrm{ and } \sigma=2$.}
\label{fig:srVSsnr}
\end{figure}
\begin{figure}
\centering\includegraphics[width=1\columnwidth]{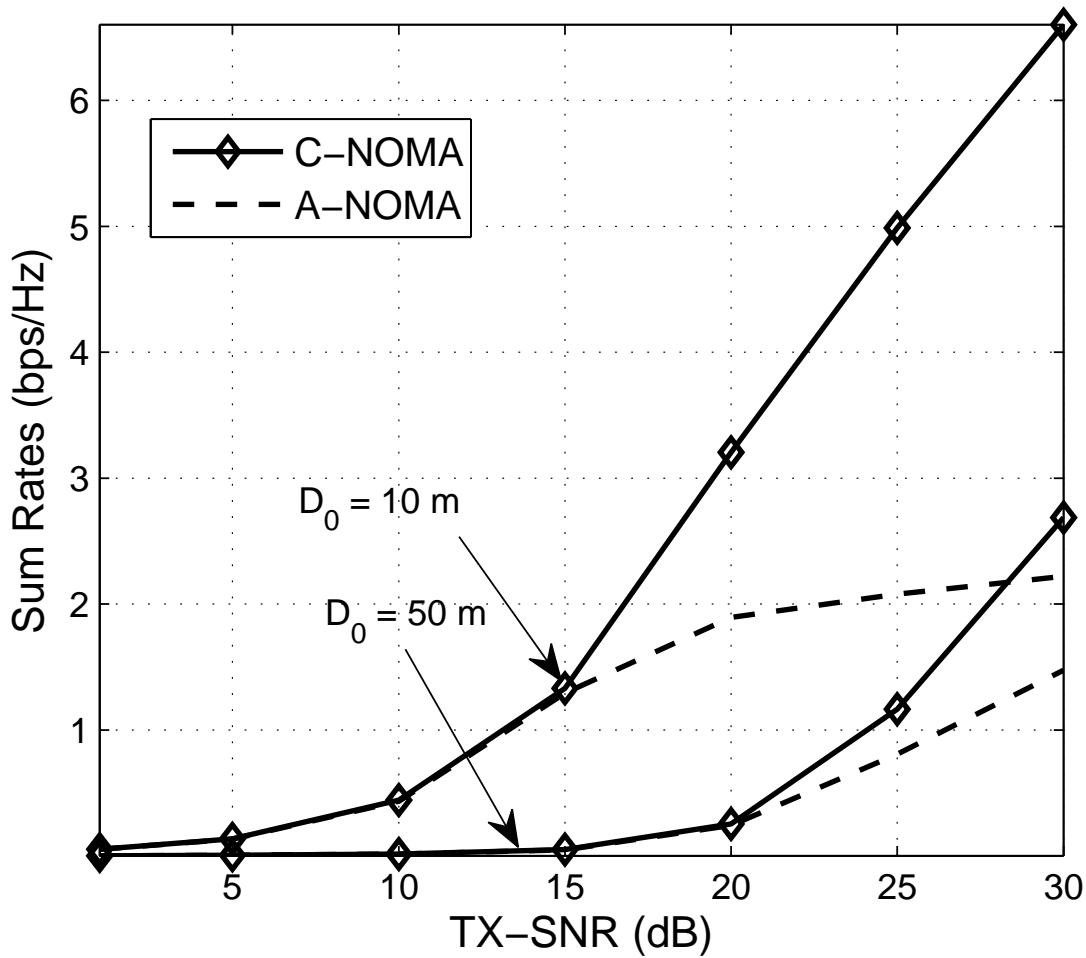}\protect\caption{Variation of average sum rates with TX-SNR for different $D_0$. $N=T=4,\sigma=1 \textrm{ and } \gamma=2.0$.}
\label{fig:srVSsnrDis}
\end{figure}
\begin{figure}
\centering\includegraphics[width=1\columnwidth]{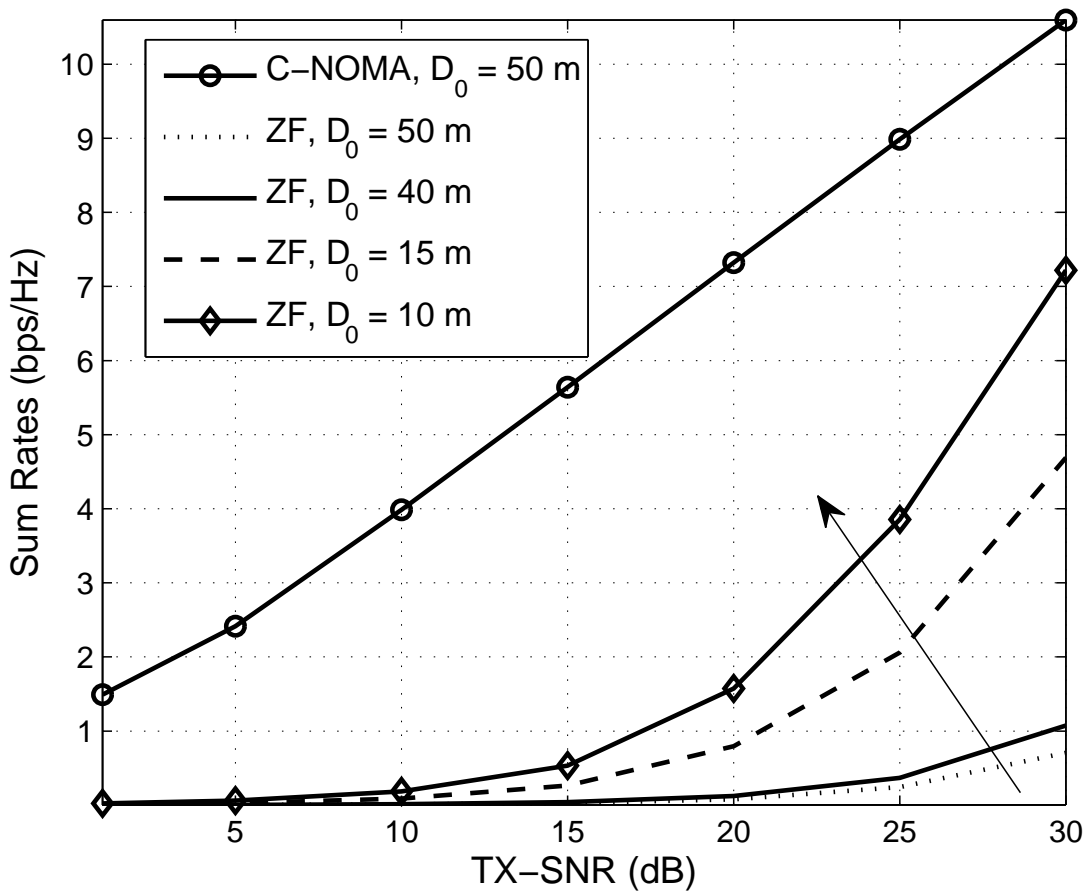}\protect\caption{The achievable sum rates as a fucntion of TX-SNR for different values of $D_0$. The parameters taken are
$N=T=4,\sigma=1 \textrm{ and } \gamma=2.0$.}
\label{fig:zfVSnomaDis}
\end{figure}
\begin{figure}
\centering\includegraphics[width=1\columnwidth]{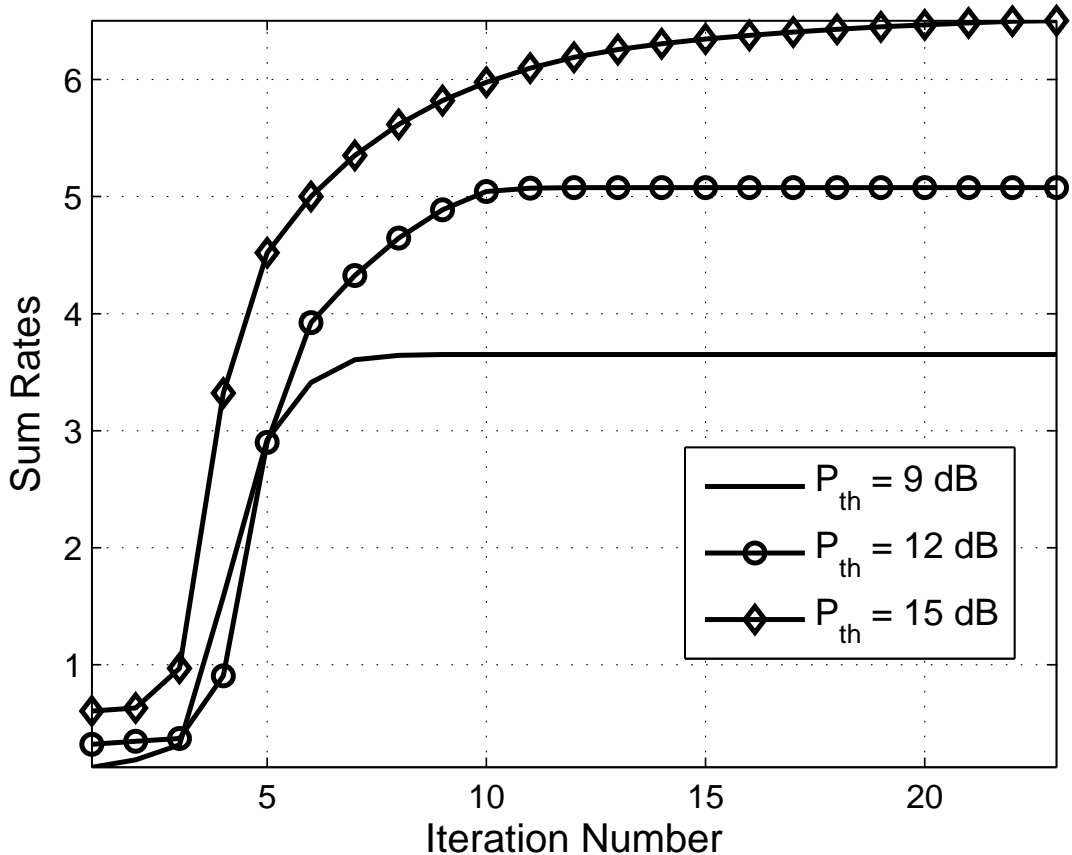}\protect\caption{Iterations required for convergence in C-NOMA approach. We take $T=N=5,\sigma=1,D_0=10 \textrm{ and }\gamma=2$. }
\label{fig:conv}
\end{figure}
\begin{figure}
\centering\includegraphics[width=1\columnwidth]{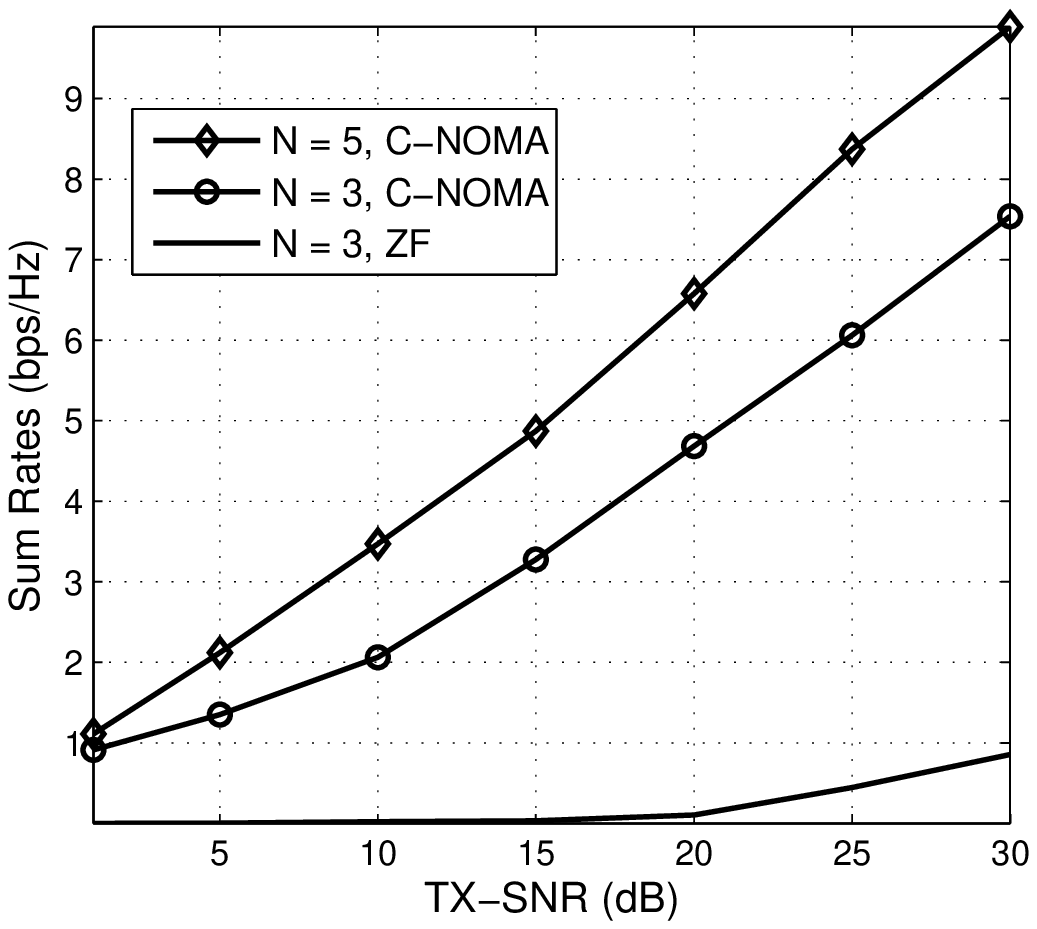}\protect\caption{The effect of TX-SNR on achievable average sum rates when $N>T$. The parameters used are $T=3,\sigma=1,\gamma=2.0 \textrm{ and }
D_0=50$.}
\label{fig:NgrT}
\end{figure}
In this section we investigate the performance of the proposed solution to the
NOMA sum rate maximization problem. For a given set of antennas $T$ and users $N$, the channels as ${\vec h}_i=\sqrt{d_i^{-\gamma}}{\vec g}_i$ are generated,
where ${\vec g}_i\sim \mathcal{CN}(0,{\vec I})$, and the distances of all users are
fixed, such that they are equally spaced between distances of $1$ and $D_0$ from the BS. It should be noted here
that in simulations the user distances are fixed and the average is taken over the fast fading component of the channel vectors. For each set of results the
values of $\gamma$ and $D_0$ are mentioned, while it is assumed that $\sigma=1$ for all users. Similarly, the transmit power is normalized with respect to noise, whose variance is
taken to be unity. For the simulations the CVX package \cite{cvx} is used.\par
In Fig. \ref{fig:srVSsnr}, we plot the average sum rates versus the transmit power for a three user system and a BS equipped with three antennas. We
take $\gamma=2,D_0=50$, and therefore,
the three users are placed at $1,25.5$ and $50$ meters from the BS, respectively. Unless specifically pointed out, $\gamma$ and $D_0$ retain the same value.
It is noted that for transmit power up to $25$ dB, the sum rates of the complete NOMA (C-NOMA) formulation and the
its approximation (A-NOMA) are equal. This observation is because of the distance effect, the ordering of the channels $\|{\vec h}_1\|_2\leq \|{\vec h}_2\|_2\leq \|{\vec h}_3\|_2$
is valid for all realizations of ${\vec g}_i$. As a consequence, $\textrm{SINR}_1^1<\min\left(\textrm{SINR}_2^1,\textrm{SINR}_3^1\right),\textrm{ and }
\textrm{SINR}_2^2<\textrm{SINR}_3^2$. Therefore, the objective function in \eqref{Rsum} matches with that in \eqref{ObjN}. Because of the wide range of multiplicative distance
factor, this observation can be attributed as a result of the Lemma \ref{chnCond}. Once the transmit power crosses a certain value ($25$ dB in our case), the
ordering of users need not to be the optimal one and hence the two curves deviate from each other. The A-NOMA approach produces better rates
because the interference free rates of the last user are boosted more compared to the C-NOMA. This comes with a degradation in the
sum rates of the $N-1$ users (excluding UE-$N$) as we will see in the next experiment. Interestingly, the competing zero-forcing (ZF) solution performs very poorly
for lower SNRs, and only produces significant sum rates, when the transmit power is sufficiently high. This poor performance of the ZF scheme can be attributed to the distance effect, which makes the
channel matrix poorly-conditioned \cite{MoliscWC}. At higher transmit SNRs this poor condition of the channel matrix is partially circumvented and hence a notable increase
in ZF rates is observed.\par
The next set of results presented in Fig. \ref{fig:srVSsnrDis} depict the average sum rates of all users excluding UE-$N$ as a function of transmit power, with $N=T=4$. Basically,
Fig. \ref{fig:srVSsnrDis} can be seen as complementing the
observations made in Fig. \ref{fig:srVSsnr}, where also at high transmit SNR A-NOMA has better total sum rates, compared to C-NOMA. It is seen that for low SNRs the curves for A-NOMA and C-NOMA
overlap. As the the transmit power is further boosted, C-NOMA outperforms A-NOMA. The reason for the equality of the rates in both techniques is the same as mentioned above.
However, at higher transmit SNR the C-NOMA provides better data rates, because of lack of optimality in the users' ordering, the beamformers of A-NOMA will not necessarily produce optimum
$\displaystyle \min_{k\leq j\leq N}(\textrm{SINR}_k^j)$ for all $k$. In addition, we have also included curves, when $D_0$ is decreased from 50 to 10 meters. It is evident that because of
the shorter distance the net effect of
distance attenuation, which orders the channels, is diminished. Hence, the gap between the graphs of C-NOMA and A-NOMA is enlarged. Nonetheless, overall higher data rates are reported in this
case because of better channel conditions for the users.\par
It can be concluded from the previous discussions, that distance plays an important role in determining the aggregate data rates of the NOMA system. Therefore, to further explore its impact we
set $N=T=4$ and plot the curves for the sum rates of C-NOMA and ZF, with $\gamma=2$. The sum rates of the ZF scheme are shown in Fig. \ref{fig:zfVSnomaDis} as the distance $D_0$ is decreased from 50 m to 10 m.
As the distance is decreased, the effect of path loss is minimized and we have better conditioned channel matrices. Therefore, the sum rates of ZF are
considerably enhanced at $D_0=10$ m.\par
In order to investigate the convergence of the proposed algorithm, we consider a downlink system with $T=5$ antennas, serving $N=5$ users. As a stopping criteria, we use
successive values of the sum rate returned by the algorithm. The algorithm exits from the main sequential iteration loop, when the difference between two
consecutive values of the sum rate is less than or equal to $10^{-2}$. With  this criterion, as shown in Fig. \ref{fig:conv}, the algorithm
converges within $25$ iterations for the three values of transit SNR shown in the figure. Moreover, as expected, with higher transmit power, we obtain better sum rate.\par
As a multiuser system is considered, the proposed approach is expected to deliver acceptable spectral efficiency when $N>T$. The results reported in Fig. \ref{fig:NgrT} show the
performance of C-NOMA, when the number of users $N$ is greater than the number of transmit antennas $T=3$. For comparison we have also included the sum rates achieved by the C-NOMA and
ZF solutions with $N=3$ users only. To obtain these
two curves, we randomly pick three users to be served with C-NOMA and ZF precoders. It is evident that with fewer users C-NOMA underperforms. Since, in this case, users are
randomly chosen, it is likely that the effective multiuser diversity \cite{YooZF} is lost and we see a downward trend in achievable data rates.
\section{Conclusion}\label{Conclusion}
In this paper, we have studied the sum rate maximization problem of a MISO downlink system based on NOMA. Specifically, we approximate the originally non-convex optimization
problem with a MM method. For the proposed algorithm, we have solved an SOCP with polynomial computational complexity in each step. For the scenarios considered, the algorithm
is numerically shown to converge within a few iterations. Furthermore, we developed a reduced complexity approximation and explore the conditions under which
it is tight. Finally, we provide an insight into the tightness of the proposed approximation. Our experimental results reveal that the NOMA has a superior performance compared to conventional orthogonal multiple access schemes. High data rates are obtained
with small transmit power. The distance attenuation has a very low impact on NOMA performance. NOMA particularly outperforms ZF when the number of users is higher than
the transmit antennas, thus making it an ideal candidate for enabling multiple access in the next generation $5$G networks.
\appendices
\section{Proof of Proposition \ref{Feast}}\label{proofFeast}
Without loss of generality, we focus on the function $f(\boldsymbol{\uptheta}_{k,k})$, its approximation $g(\boldsymbol{\uptheta}_{k,k},\boldsymbol{\uptheta}_{k,k}^t)$ and the
constraint in which it appears. The
same arguments will be applicable to all non-convex functions, their convex minorants and the respective constraints. Therefore, it holds that
\begin{align}
0.25(\bar{w}_k+r_k)^2-\bar{w}_k-g(\bar{w}_k,r_k,\bar{w}^t_k,r^t_k)\leq g(\boldsymbol{\uptheta}_{k,k},\boldsymbol{\uptheta}_{k,k}^t),\label{RprobaTR2NCvx5FrPR}
\end{align}
where $g(\bar{w}_k,r_k,\bar{w}^t_k,r^t_k)\triangleq0.25\left[(\bar{w}_k^t-r_k^t)^2+2(\bar{w}^t_k-r^t_k)\{\bar{w}_k-\bar{w}^t_k-
r_k+r_k^t\}\right]$ is the approximation of the original function $(\bar{w}_k-r_k)^2$. Note, that this constraint
is a convex approximation of that in \eqref{RprobaTR2NCvx1}. Now, let us assume that the tuple $(\bar{w}_k^t,r_k^t,\boldsymbol{\uptheta}_{k,k}^t)$ is feasible to
\eqref{RprobaTR2NCvx1}. Clearly, the same point also satisfies \eqref{RprobaTR2NCvx5FrPR} as a consequence of \eqref{propgTh2}. Since
$g(\bar{w}_k,r_k,\bar{w}^t_k,r^t_k)\leq(\bar{w}_k-r_k)^2$
and $f(\boldsymbol{\uptheta}_{k,k})\geq g(\boldsymbol{\uptheta}_{k,k},\boldsymbol{\uptheta}_{k,k}^t)$, it follows that
\begin{align}
0.25(\bar{w}_k+r_k)^2-&\bar{w}_k-0.25(\bar{w}_k-r_k)^2-f(\boldsymbol{\uptheta}_{k,k})\\& \leq 0.25(\bar{w}_k+r_k)^2-\bar{w}_k-
g(\bar{w}_k,r_k,\bar{w}^t_k,r^t_k)-g(\boldsymbol{\uptheta}_{k,k},\boldsymbol{\uptheta}_{k,k}^t).\label{seQProFes1}
\end{align}
Hence, $(\bar{w}_k^{t+1},r_k^{t+1},\boldsymbol{\uptheta}_{k,k}^{t+1})$ should satisfy \eqref{RprobaTR2NCvx1} because
\begin{align}
0.25(\bar{w}_k^{t+1}+&r_k^{t+1})^2-\bar{w}_k^{t+1}-0.25(\bar{w}_k^{t+1}-r_k^{t+1})^2-f(\boldsymbol{\uptheta}_{k,k}^{t+1})\\& \leq 0.25(\bar{w}_k^{t+1}+r_k^{t+1})^2-\bar{w}_k^{t+1}-
g(\bar{w}_k^{t+1},r_k^{t+1},\bar{w}^{t}_k,r^{t}_k)-g(\boldsymbol{\uptheta}_{k,k}^{t+1},\boldsymbol{\uptheta}_{k,k}^{t})\leq 0.\label{seQProFes2}
\end{align}
The above conclusion holds for all $k$ and $\{\mathcal{V}_t\}_{t\geq 0}$, as the algorithm was initialized with $\boldsymbol{\Uplambda}^{(0)}\in\mathcal{F}_0$.
\section{Proof of Proposition \ref{Ascent}}\label{proofAscent}
In order to prove this proposition, we note that $\mathcal{F}_{t+1}\supseteq \mathcal{F}_t$. From \eqref{serInq1} it is
clear that the surrogate functions used in place of non-convex terms are non-decreasing with iteration number i.e., $SF^{t+1}
\geq SF^{t}$, where $SF$ is a generic representation of these functions used in the paper and is valid for all of them.
Therefore, $\mathcal{F}_{t+1}\supseteq \mathcal{F}_t$, is an immediate consequence, and the statement in Proportion \ref{Ascent} follows.
Hence, $\{\mathcal{O}_t\}_{t\geq 0}$ is non-decreasing, and possibly converges to positive infinity.
\section{Proof of Proposition \ref{KKT-conv}}\label{proofKKT}
The following assumptions are made before outlining the arguments.
\newtheorem{assumption}{Assumption}
\begin{assumption}
We assume that as $t\rightarrow \infty$, the sequence of variables $\{\mathcal{V}_t\}_{t\geq 0}$ generated by the algorithm in Table \ref{ALG}
converges to a value $\mathcal{V}^{*}$.
\end{assumption}
\begin{assumption}
The constraints in the approximate problem \eqref{TthRun0} or \eqref{simpTthRun0} are
qualified at the accumulation point.
\end{assumption}
Without explicitly mentioning the constraints, we use abstract notation to prove the claim made in Proposition \ref{KKT-conv}. First let us give a generic representation to all
convex constraints in \eqref{TthRun0} as $\mathcal{C}_a\mathcal{(V)}_s\leq 0, a=1,\ldots,L_1$, where $\mathcal{(V)}_s$ denotes the subset of $\mathcal{V}_t$ containing the corresponding variables
that appear in these constraints. Similarly, let us define as $\mathcal{C}_b^t\mathcal{(V)}_p\leq 0, b=L_1+1,\ldots,L_2$ the constraints obtained by approximating the non-convex
functions with convex minorants in \eqref{TthRun0}, and $\mathcal{(V)}_p\subseteq \mathcal{V}_t$. Let $\eta_a^*,\bar{\eta}_b^*\in\mathbb{R}_+ \textrm{ for all } a,b$,
denote the dual variables at convergence. The KKT conditions of the
problem in \eqref{TthRun0} at $\mathcal{(V^*)}_s,\mathcal{(V^*)}_p$ then read as
\begin{align}
&\nabla {\vec r}^*+\sum_{a=1}^{L_1}\eta_a^*\nabla \mathcal{C}_a\mathcal{(V^*)}_s+\sum_{b=L_1+1}^{L_2}\bar{\eta}_b^*\nabla \mathcal{C}_b^t\mathcal{(V^*)}_p=0\label{KKTcd1}\\&
\eta_a^* \mathcal{C}_a\mathcal{(V^*)}_s=0,\:a=1,\ldots,L_1,\qquad\bar{\eta}_b^* \mathcal{C}_b^t\mathcal{(V^*)}_p=0,\:b=L_1+1,\ldots,L_2\label{KKTcd2}.
\end{align}
Since all convex minorants satisfy the properties in \eqref{propgTh0}, it is easy to
conclude that the KKT conditions given above will reduce to those of the problem in \eqref{TthRun0}. Similar conclusion also holds for the
simplified problem in \eqref{simpTthRun0}.
\section{Proof of Lemma \ref{chnCond}}\label{proofchnCond}
For \eqref{ObjN} to be valid for all $1\leq k\leq N-1$, it holds that
\begin{align}\textrm{SINR}_k^k<\min_{i\in[k+1,N]}\textrm{SINR}_i^k.\end{align} For an arbitrary $k\in[1,N-1]$ and $i=n$, let us consider the following inequality,
\begin{align}
\frac{|{\vec h}_{n}\herm{\vec w}_k|^2}{\sum_{m=k+1}^N|{\vec h}_{n}\herm{\vec w}_m|^2+\sigma_{n}^2}> \frac{|{\vec h}_k\herm{\vec w}_k|^2}
{\sum_{m=k+1}^N|{\vec h}_k\herm{\vec w}_m|^2+\sigma_{k}^2},\label{chnCond1}
\end{align}
where we have assumed that the noise variances at the $n^{\textrm{th}}$ and the $k^{\textrm{th}}$ nodes are
$\sigma_{n}^2$ and $\sigma_{k}^2$, respectively.
By substituting the assumptions made in the lemma, ${\vec h}_{n}=c_nc_{n-1}\ldots c_{k+1}{\vec h}_k\triangleq c_n^{k+1}{\vec h}_k$, where
$k+1\leq n \leq N$.
After some simple manipulations
\begin{align}
{\sum_{m=k+1}^N|{\vec h}_k\herm{\vec w}_m|^2+\sigma_{k}^2}> \sum_{m=k+1}^N|{\vec h}_{k}\herm{\vec w}_m|^2+\sigma_{n}^2/|c_n^{k+1}|^2\Leftrightarrow
|c_n^{k+1}|>\frac{\sigma_{n}}{\sigma_{k}}.\label{chnCond2}
\end{align}
Now, if $\sigma_{n}=\sigma_k$, using the condition on $|c_n^{k+1}|$, we obtain $\|{\vec h}_{n}\|_2> \|{\vec h}_k\|_2$ for all $n$. 
Repeating the same argument for all $k$, the required proof follows.
\section{Proof of Lemma \ref{chnCondanother}}\label{proofchnCondanother}
The $\textrm{SINR}_i^k$, $i\geq k$, can be written as
\begin{align}
&\textrm{SINR}_i^k=\frac{|{\vec h}_i\herm{\vec w}_k|^2}{\sum_{m=k+1}^N|{\vec h}_i\herm{\vec w}_m|^2+\sigma^2}.
\end{align}
If a random unitary matrix is used for precoding, $|{\vec h}_i\herm{\vec w}_k|^2$ is still complex Gaussian distributed, since a unitary transformation
of Gaussian vectors is still complex Gaussian distributed. In addition, $|{\vec h}_i\herm{\vec w}_k|^2$ and $|{\vec h}_i\herm{\vec w}_l|^2$, $k\neq l$ are
independent. Define $x_{ik}\triangleq |{\vec h}_i\herm{\vec w}_k|^2$ and $y_{ik}\triangleq \sum_{m=k+1}^N|{\vec h}_i\herm{\vec w}_m|^2$.  Therefore $x_{ik}$ is an
exponentially distributed random variable, with $\lambda_{i}\triangleq d_i^\gamma$, i.e., $f_{x_{ik}}(x)=\lambda_ie^{-\lambda_ix}$. Similarly, $y_{ik}$ follows
the Chi-square distribution, i.e.,  \begin{align}f_{y_{ik}}(y)=\frac{\lambda_i^{N-k}y^{(N-k-1)}}{(N-k-1)!}e^{-\lambda_iy}.\end{align} Consequently the cumulative distribution function
of $\textrm{SINR}_i^k$ can be calculated from the following
\begin{align}
\mathrm{Pr}\left(\textrm{SINR}_i^k\leq \theta \right)&=\mathrm{Pr}\left( \frac{x_{ik}}{y_{ik}+\sigma^2}\leq \theta \right)\\
&= \int_{0}^{\infty}\left( 1-e^{-\lambda_i\theta \left(y+\sigma^2\right)}\right)f_{y_{ik}}(y)dy\\
&= 1- \frac{e^{-\lambda_i\theta \sigma^2}}{(N-k-1)!}\int_{0}^{\infty} e^{-(1+\theta)\lambda_i y} (\lambda_i y)^{(N-k-1)} d\lambda_i y.
\end{align}
Applying \cite[Eq. (3.351.3)]{intg}, the pdf of $\textrm{SINR}_i^k$ can be obtained as follows:
\begin{align}
F_{\textrm{SINR}_i^k}(z)&=1-\frac{e^{-\lambda_i \sigma^2 z}}{(1+z)^{N-k}}.
\end{align}
Again, following the unitary transformation of Gaussian variables, the desired probability can be evaluated as
\begin{align}
\mathrm{Pr}\left(\textrm{SINR}_i^k>\textrm{SINR}_j^k\right)&=\int^{\infty}_{0} \left(1-\frac{e^{-\lambda_j \sigma^2 z}}{  (1+z)^{N-k}}\right) f_{\textrm{SINR}_i^k}(z)dz\\
&=1-\int^{\infty}_{0}   \left(\frac{\lambda_i\sigma^2 e^{-(\lambda_i+\lambda_j) \sigma^2 z}}{(1+z)^{2(N-k)}}+\frac{(N-k)e^{-(\lambda_i+\lambda_j) \sigma^2 z}}{(1+z)^{2(N-k)+1}}\right)dz.
\end{align}
Applying \cite[Eq. (3.351.4)]{intg}, the above probability can be expressed as
\begin{align}\label{zhi_probab}
\mathrm{Pr}\left(\textrm{SINR}_i^k>\textrm{SINR}_j^k\right)
&=1-e^{(\lambda_i+\lambda_j)\sigma^2} \lambda_i\sigma^2  \psi\left((\lambda_i+\lambda_j)\sigma^2,2(N-k)\right)\\
\notag&- e^{(\lambda_i+\lambda_j) \sigma^2}(N-k)
\psi\left((\lambda_i+\lambda_j) \sigma^2,2(N-k)+1\right),
\end{align}
and the proof is completed.

\bibliographystyle{IEEEtran}
\bibliography{IEEEabrv,NOMA_Spec_Eff_Max}

\end{document}